\documentclass[reprint,amsmath,amssymb,aps,preprintnumbers,nofootinbib]{revtex4-2}

\usepackage{graphicx}
\usepackage[colorlinks=true, linkcolor=blue, citecolor=blue, urlcolor=blue]{hyperref}  
\usepackage{mhchem}
\usepackage{array}
\usepackage{pdfpages}
\usepackage{pgffor}

\makeatletter
\AtBeginDocument{\let\LS@rot\@undefined}
\makeatother

\def\supplementfilename{supplement.pdf}
\pdfximage{\supplementfilename}
\def\numbersupplementpages{\the\pdflastximagepages}

\newif\ifarXiv
\arXivtrue   

\begin{document}

\title{Energy-Resolved Spin and Parity Distributions of Fission Fragments}

\author{Petar Marevi\'c}
\email{Corresponding author: pmarevic@phy.hr}
\affiliation{University of Zagreb, Faculty of Science, Department of Physics,
HR-10000 Zagreb, Croatia}

\author{Nicolas Schunck}
\affiliation{Nuclear data and Theory Group, Nuclear 
and Chemical Science Division, Lawrence Livermore National Laboratory, 
California, USA 94550}

\author{Antonio Bjel\v{c}i\'c}
\affiliation{Applied Mathematics and Computational Research Division,
Lawrence Berkeley National Laboratory, California, USA 94720}

\author{Guillaume Scamps}
\affiliation{Universit\'e de Toulouse, CNRS/IN2P3, L2IT, Toulouse, France}

\begin{abstract}
Fission fragments are born rotating rapidly, but the evolution of this spin with 
excitation energy and its impact on fission observables remain largely unknown. 
Combining state-of-the-art microscopic and statistical reaction theory, we 
simulate the full fission process -- from the initial state, through scission, 
to fragment decay -- and report the first microscopic, energy-resolved spin and 
parity distributions of the fragments produced in $^{239}$Pu(n,f). From thermal 
energy to $E_n = 10$~MeV, the spin of individual fragments evolves non-monotonically,
tracking their deformation. The parity content, strongly non-statistical at 
thermal energy, remains correlated with spin throughout this range. For the first 
time, we quantify the impact of microscopic spin--parity distributions on prompt 
photon multiplicities, demonstrating the predictive potential of parameter-free microscopic 
inputs.
\end{abstract}

\date{\today}

\maketitle


Nuclear fission transforms a heavy nucleus carrying little or no angular 
momentum into two rapidly spinning fragments. More than five decades after 
this observation \cite{wilhelmy1972}, the mechanism responsible for generating 
such large angular momenta from a near-spinless parent remains a central open 
question in low-energy nuclear physics \cite{schunck2022,bender2020}. The 
fragment spins shape the emission of prompt $\gamma$ rays, which serve as
signatures of actinide production in neutron star mergers \cite{eichler2015}, 
underpin active interrogation techniques \cite{bendahan2020}, and contribute to the 
heat deposition in nuclear reactors \cite{rimpault2012}. Interest in fragment 
spins was recently renewed by measurements across several actinides that 
revealed a universal sawtooth dependence of the mean spin on fragment mass 
\cite{wilson2021}, followed by a series of microscopic \cite{bulgac2021,
marevic2021,scamps2023b,scamps2025,marevic2026,li2026} and statistical 
\cite{randrup2021,stetcu2021,randrup2022a,randrup2022b,dossing2024} studies. 
Most of these studies, however, were carried out at a single, low excitation 
energy, leaving the dependence of the fragment spins on the excitation energy 
of the fissioning nucleus largely unexplored.

In induced fission, this energy is determined by the kinetic energy of the 
incident neutron, and its impact on the fragment spin has implications for 
both fundamental and applied physics. Competing generation mechanisms predict 
distinct spin-energy dependencies, and mapping this dependence can help constrain 
the underlying mechanism \cite{giha2023}. Fragment spin distributions are also 
an essential input to the statistical reaction theory codes \cite{talou2021,
verbeke2018,litaize2015,ormand2021monte,okumura2018,koning2023talys} used to 
simulate fragment decay, and therefore enter the nuclear data evaluation process 
supporting a range of critical applications. This input, however, is poorly constrained 
experimentally, as fragment spins can only be inferred from the subsequent decay 
radiation. Several $\gamma$-multiplicity measurements show that the number of 
emitted photons grows with neutron energy \cite{qi2018,laborie2018,giha2023}. 
In $^{239}$Pu(n,f), this growth is approximately linear below the second-chance-fission 
threshold, evidence of a positive spin-energy correlation \cite{giha2023}, 
though such measurements infer the spin only from the total photon count. A 
$\gamma$-ray spectroscopy campaign resolving individual transitions over a broad 
range of neutron energies is planned for 2026 at the NFS facility of GANIL 
\cite{wilson_talk}. Fragment parity is even less constrained, and decay codes 
generally assume an equiprobable partition between positive and negative parity.

Predicting the spin and parity of fission fragments across a range of neutron energies 
therefore requires a microscopic theory, in which fission dynamics emerges 
self-consistently from the nucleonic degrees of freedom, without parameters 
adjusted to fission observables \cite{schunck2022,schunck2016}. Such 
calculations, however, remain scarce and limited. An initial study reporting 
an increase of the  mean spin with neutron energy considered only the most 
likely fragments, without resolving individual fragments, and did not 
specify the excitation mechanism \cite{bulgac2021}. The work of 
Ref.~\cite{marevic2026} was limited to cold fragments and adiabatic dynamics, and 
could therefore establish only a lower bound on the effect. The most recent 
results on the dependence of fragment spin on the fissioning-system excitation 
energy and spin were restricted to a single fragmentation and to mean spins 
\cite{qiang2026}. Furthermore, the parity content of the fragments has only 
very recently been addressed microscopically, and only at thermal energy 
\cite{scamps2026}. Microscopic predictions of spin and parity distributions 
of fission fragments, resolved in incident neutron energy, are therefore still 
lacking and their impact on prompt fission observables is unknown.


In this Letter, we report microscopic, energy-resolved spin and parity 
distributions of the primary fragments produced in $^{239}$Pu(n,f) for 
incident neutron energies from thermal to $10$ MeV. Fission dynamics 
is simulated using a combination of the time-dependent Hartree-Fock-Bogoliubov 
(TDHFB) theory \cite{simenel2025,Bulgac2019b} and the time-dependent 
generator coordinate method (TDGCM) \cite{regnier2016,verriere2020}. 
TDHFB provides the most complete microscopic description of beyond-saddle dynamics 
currently feasible. We use the \textsc{axialhohfb} program~\cite{bjelcic2026}, 
based on an optimized axially-symmetric harmonic oscillator basis, which enables a 
well-converged description of fission dynamics at unprecedentedly low 
computational cost. TDGCM, as implemented in the \textsc{felix} program 
\cite{regnier2018}, is the tool of choice for describing the adiabatic phase 
of the dynamics, before the saddle point, and is used here to determine 
the population of each TDHFB trajectory. All calculations employ the SkM* 
\cite{bartel1982} Skyrme energy density functional with a mixed pairing 
interaction \cite{dobaczewski2002}.

The initial configurations for the TDHFB evolution, corresponding to 
$^{240}$Pu$^*$ formed in thermal-neutron fission of $^{239}$Pu, are 
prepared through HFB calculations with constraints imposed on the 
expectation values of axial quadrupole ($q_{20}$) and octupole ($q_{30}$) 
moments. We generate a set of ten configurations beyond the saddle 
point, with energies $E \in [6.35, 6.65]$~MeV above the ground state, 
matching the excitation energy of $^{240}$Pu$^*$ at the one-neutron 
separation energy $S_n \approx 6.5$~MeV. Each configuration is evolved 
in time until the fragments are fully separated. The resulting trajectories 
yield a broad range of average fragment masses, spanning the asymmetric 
fragmentations that dominate thermal fission.

The effect of nuclear excitation is simulated by applying an instantaneous 
quadrupole boost to the initial configurations \cite{goddard2016,simenel2025}. 
Each configuration is thereby excited to an energy $E^*$ above the ground 
state, with the equivalent incident neutron energy estimated as $E_n = E^* - S_n$. 
We boost the full set of configurations to energies corresponding to 
$E_n = 1, 3, 5, 7, 10$~MeV. The deposited collective energy is rapidly 
(within $t < 1$~zs) transferred into the intrinsic degrees of freedom 
\cite{SM,goddard2016}, as the initially coherent nucleon current thermalizes 
into incoherent single-particle excitations, leaving the system with little 
memory of the excitation mechanism. Since the descent to scission is itself 
driven by a large increase in $q_{20}$, a quadrupole boost couples this 
energy directly to the fission motion, providing a physically motivated means 
of imparting excitation energy to the system.

\begin{figure}[!t]
    \centering    
    \includegraphics[width=0.95\linewidth]{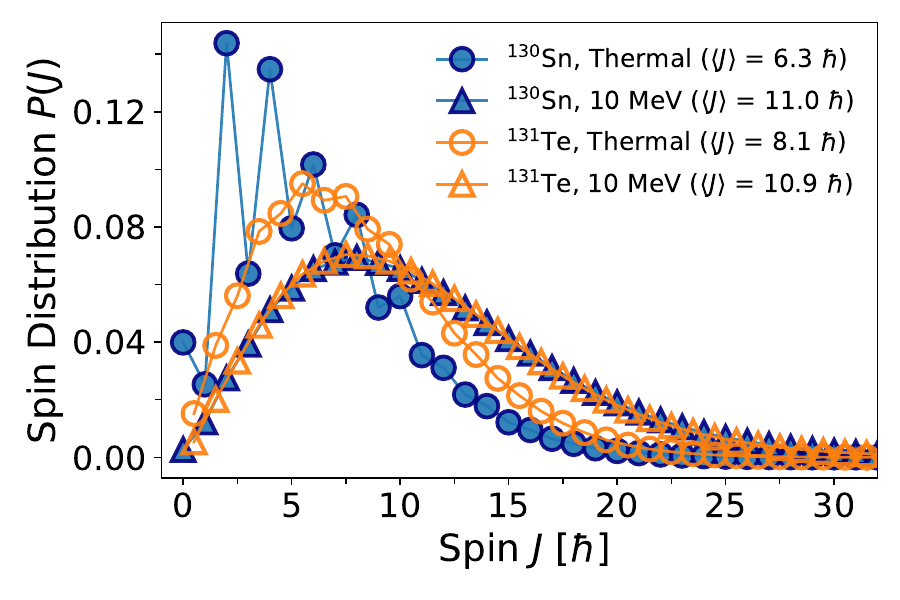}
    \caption{Spin distributions of \ce{^{130}_{50}Sn} and \ce{^{131}_{52}Te} 
    primary fragments produced in thermal and $E_n = 10$~MeV fission of 
    $^{239}$Pu.}
    \label{fig:illustrative_distribution}
\end{figure}

For fully separated fragments on all sixty trajectories, we perform a joint, 
seven-dimensional projection on fragment angular momentum $J$ and its 
fission-axis component $K$ \cite{scamps2023b}, parity $\pi$ \cite{scamps2026}, 
the fragment neutron and proton numbers $N$ and $Z$ \cite{simenel2010,
verriere2019}, and the total neutron and proton number \cite{verriere2019}. 
Projections are performed using the Pfaffian method \cite{robledo2009,
bertsch2012,scamps2013,wimmer2012}, in the rest frame of each fragment. 
This resolves each trajectory into a distribution of fragmentations, so that
a large set of $(N, Z)$ is accessible from a limited number of dynamical paths. 
Each trajectory is weighted by the collective flux reaching its initial 
configuration on the $(q_{20}, q_{30})$ potential energy surface, as determined 
by TDGCM, while neglecting interference between trajectories.  

\begin{figure*}[!ht]
    \centering    
    \includegraphics[width=0.99\linewidth]{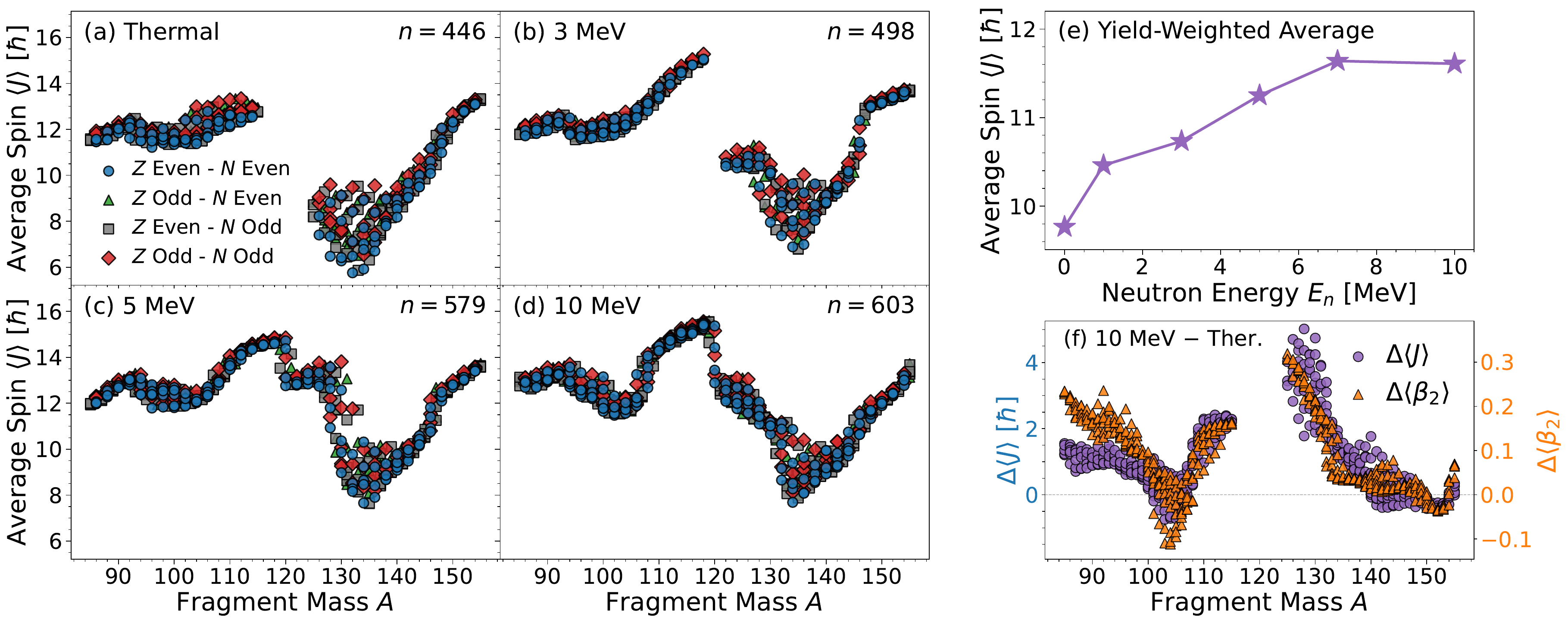}
    \caption{(a)--(d) Average spin of fission fragments as a 
    function of their mass, for thermal and $E_n = 3, 5, 10$~MeV 
    fission of $^{239}$Pu. The fragments are partitioned according 
    to whether $Z$ and $N$ are even or odd, and $n$ denotes the number 
    of predicted fragments. (e) Yield-weighted average spin
    as a function of $E_n$. (f) The difference of mean spin 
    $\Delta \langle J \rangle$ (left axis) and mean quadrupole deformation
    parameter $\Delta \langle \beta_2 \rangle$ (right axis) between 
    $E_n = 10$~MeV and thermal fission.}
    \label{fig:meanJ_vs_A}
\end{figure*}


The obtained spin distributions cover fragments of all combinations of 
proton and neutron number parity, with integer (half-integer) spins for 
even (odd) nuclei, and are additionally resolved in $\pi$ and $K$. 
Figure~\ref{fig:illustrative_distribution} shows illustrative spin 
distributions for two nuclides near the $N=82$, $Z=50$ double shell 
closure. When produced in thermal fission, the even-even \ce{^{130}_{50}Sn} 
nucleus has a large $J = 0$ component and a marked staggering pattern 
with maxima at even $J$, reflecting its quadrupole and weak octupole 
deformation. The odd-neutron \ce{^{131}_{52}Te} nucleus, in contrast, 
has a significantly smoother distribution already at thermal energy, 
carrying about two more units of spin on average. However, at 
$E_n = 10$~MeV the two distributions become nearly indistinguishable 
and are governed by their almost identical bulk properties. Such 
distributions are obtained for every fragment at each $E_n$; in the 
following we characterize the systematics through their mean values.

Figures~\ref{fig:meanJ_vs_A}(a)--(d) show the average spin of the 
fragments as a function of their mass for thermal and $E_n = 3, 5, 
10$~MeV fission. We obtain $446$ fragments at thermal energy, with 
a population probability above $10^{-6}$, distributed almost evenly 
among the four number-parity classes: $112$ even-even, $108$ odd-even, 
$118$ even-odd, and $108$ odd-odd. They form a characteristic sawtooth 
pattern, with a pronounced dip in the doubly-magic $^{132}$Sn region 
and a gap near symmetry that reflects the negligible probability of 
such fragmentations at low excitation. With increasing $E_n$ the number 
of predicted fragments grows steadily, reaching $603$ for $E_n = 10$~MeV, 
and the symmetric region becomes populated. Overall, our calculations 
predict that the sawtooth pattern persists up to at least $E_n = 10$~MeV, 
for the first-chance fission of $^{240}$Pu$^*$ considered here.

Figure~\ref{fig:meanJ_vs_A}(e) shows the average spin over all fragments, 
weighted by the calculated yields, as a function of $E_n$. It increases 
by nearly $2\,\hbar$ between thermal and $E_n = 7$~MeV, consistent with 
the experimentally inferred positive spin--energy correlation \cite{giha2023}. 
However, the increase is not uniform: it is steepest within the first MeV 
above thermal fission, rises approximately linearly between $1$ and $7$~MeV, 
and saturates thereafter. This initial steep rise occurs at energies below 
those covered by existing multiplicity measurements. These results indicate 
that the dependence may not be uniformly linear across the full energy range
-- a possibility that future experiments specifically plan to probe \cite{giha2023}.

We find that this behavior originates in the geometry of the nucleus at 
scission. The shape elongation at which the neck ruptures follows the same 
energy dependence, increasing up to $E_n = 7$~MeV and leveling off thereafter. 
The average spin is in fact a nearly linear function of the mean elongation 
at scission, and is better predicted by it than by the incident energy 
\cite{SM}. The connection between deformation and spin also holds for 
individual fragments. Figure~\ref{fig:meanJ_vs_A}(f) shows the change in mean 
spin and mean quadrupole deformation between $E_n = 10$~MeV and thermal fission, 
for all fragments appearing at both energies. Most fragments acquire spin with 
$E_n$, by up to $2.5\,\hbar$ in the light group and up to $5\,\hbar$ in the 
heavy group. In contrast, the spin of some fragments with $A \in [100, 110]$ 
and $A \in [140, 150]$ remains constant or even decreases. The two quantities 
are strongly correlated across the entire mass range, including the two regions
where both become negative \cite{SM}. 

The effect of nuclear elongation at scission thus governs the fragment spins 
at both levels: it sets both how the average spin grows with incident energy, 
and, through the deformation each fragment inherits, the distribution of 
that growth across fragment mass. Resolving this mass dependence, as planned 
at GANIL \cite{wilson_talk}, would test the prediction: the pronounced increase 
near $A \approx 130$ is the most readily accessible feature, while a reversal 
of this trend elsewhere would confirm that this increase is not universal.
A direct comparison will, however, require accounting for prompt  emissions, 
which shift the fragment mass and remove part of the spin \cite{stetcu2021}.

The present model is also able to predict the parity content of the fragments. 
Statistical decay models typically assume an equiprobable partition between
positive and negative parities, on the expectation that comparable numbers of 
levels of both parities are available at sufficiently high energies 
\cite{talou2021}. Recent microscopic calculations suggest that this 
assumption breaks down at thermal energy \cite{scamps2026}, but its validity 
at higher incident energies has not been examined.

\begin{figure}[!t]
    \centering    
    \includegraphics[width=0.99\linewidth]{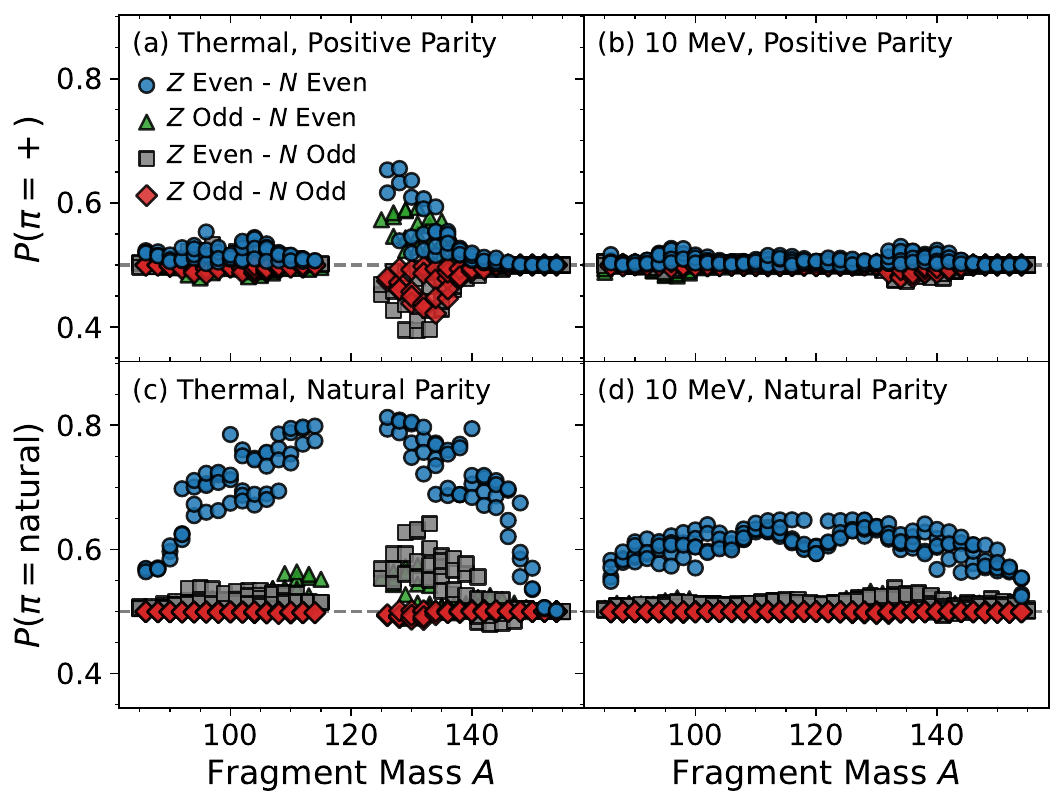}
    \caption{Parity content of fission fragments as a function
    of fragment mass, for thermal and $E_n = 10$~MeV fission.
    (a)--(b) Probability of positive parity.
    (c)--(d) Probability of natural parity. In all panels the 
    fragments are partitioned according to whether $Z$ and $N$ 
    are even or odd, and the dashed lines mark the equiprobable 
    limit.}
    \label{fig:parity}
\end{figure}

Figures \ref{fig:parity}(a)--(b) show the probability of positive parity as 
a function of fragment mass, for thermal and $10$~MeV fission. At thermal 
energy the distributions depend systematically on the number parity of the 
fragment and deviate significantly from the equiprobable value, consistent 
with Ref.~\cite{scamps2026}.  The deviations are of opposite sign in the 
even-even and odd-odd fragments, so that they largely cancel if averaged over
the whole distribution. By $E_n = 10$~MeV these deviations have largely 
disappeared. Our calculations therefore suggest that the assumption of 
equipartition between positive and negative parity is poorest at thermal 
energy and becomes adequate at a few MeV.

Statistical decay models, however, assume an equal population of both 
parities at every spin, which is a stronger condition. It is better captured 
by a complementary classification relating parity to spin: states with 
$\pi = (-1)^J$ for integer $J$, and $\pi = (-1)^{J-1/2}$ for half-integer 
$J$, are termed natural, and the remainder unnatural. For some fragments the 
two parities may be equally populated overall while the parity alternates 
with spin, so that the joint distribution does not factorize into independent 
spin and parity distributions.

Figures~\ref{fig:parity}(c)--(d) show the  probability of natural parity for 
thermal and $10$~MeV fission. At thermal energy the even-even fragments, 
have the largest fraction of natural parity, up to $0.8$. This fraction 
decreases markedly with $E_n$, reaching roughly $0.6$ for all even-even 
fragments at $10$~MeV. Fragments with an odd number of neutrons or protons 
show a slight preference for natural parity, which is fully erased by 
$E_n = 10$~MeV. The odd-odd fragments exhibit no $E_n$ dependence, their 
parity content already being close to the statistical limit at thermal 
energy. These trends follow the number of pairs available to be broken 
in each class, consistent with a progressive quenching of pairing 
correlations with excitation energy. Even at $10$~MeV, however, the 
even-even fragments remain well away from the uncorrelated value of 
$0.5$, so the factorized form assumed in decay models is not recovered 
over the energy range considered here.

Spin-parity distributions have a substantial impact on the emission of 
prompt $\gamma$-rays by the excited fragments, and are thus an essential 
ingredient of the statistical-reaction-theory codes \cite{talou2021,
verbeke2018,litaize2015,ormand2021monte,okumura2018,koning2023talys}. 
Currently, these codes obtain spin distributions from a phenomenological 
formula whose parameters are fit to reproduce measured observables such 
as the average photon multiplicity $\bar{N}_\gamma$ \cite{talou2021}. 
This limits the predictive power of the approach, especially in regimes 
where experimental data are scarce \cite{stetcu2020evaluation}.  
Moreover, equiprobable partition of parities is always assumed, even 
though the validity and impact of this hypothesis have never been tested.

Here, we compare $\bar{N}_\gamma$ predictions obtained using phenomenological 
and microscopic spin-parity distributions. To this end, we employ the 
\textsc{feta} framework \cite{schunck2026fission}, based on the Hauser-Feshbach 
reaction code \textsc{yahfc} \cite{ormand2021monte}, and compute 
$100,000$ Monte Carlo events. Four different prescriptions for the fragment 
spin distribution are considered: the parametrized phenomenological model of 
Ref.~\cite{talou2021}, the static HFB distributions \cite{marevic2026,
dataset2025}, and the present TDHFB results with ($P_{J\pi}$) and without 
($P_J$) parity restoration. All prescriptions therefore assume an independent, 
equipartitioned parity distribution, except for the TDHFB result with parity 
restoration. The remaining ingredients of the simulation are kept fixed \cite{SM}.

For thermal neutrons, the average photon multiplicity is relatively well 
constrained experimentally: $7.35 \pm 0.12$ \cite{gatera2017} and $7.23 
\pm 0.22$ \cite{verbinski1973}. With the current numerical setting, the 
phenomenological model gives $\bar{N}_{\gamma} = 7.36$. Static HFB 
distributions describe cold fragments and therefore underestimate spins, 
giving $\bar{N}_{\gamma} = 5.81$. Including the intrinsic excitation effects 
within TDHFB raises the multiplicity to $\bar{N}_{\gamma}=7.65$, while 
accounting for the microscopic parity distribution raises it further, 
significantly exceeding the measured value ($\bar{N}_{\gamma}=8.93$). 
This is accompanied by a drop in neutron multiplicity, from $2.78$ 
(phenomenological) to $2.67$ (full microscopic input), reflecting the 
stronger $n$--$\gamma$ competition favored by larger spins. Since the 
parameters of the decay model were adjusted with a parametrized spin 
distribution, this suggests some of them, e.g., the excitation-energy 
partition between the fragments, may need recalibrating once microscopic 
spin-parity input is used.

\begin{figure}[t]
    \centering    
    \includegraphics[width=0.97\linewidth]{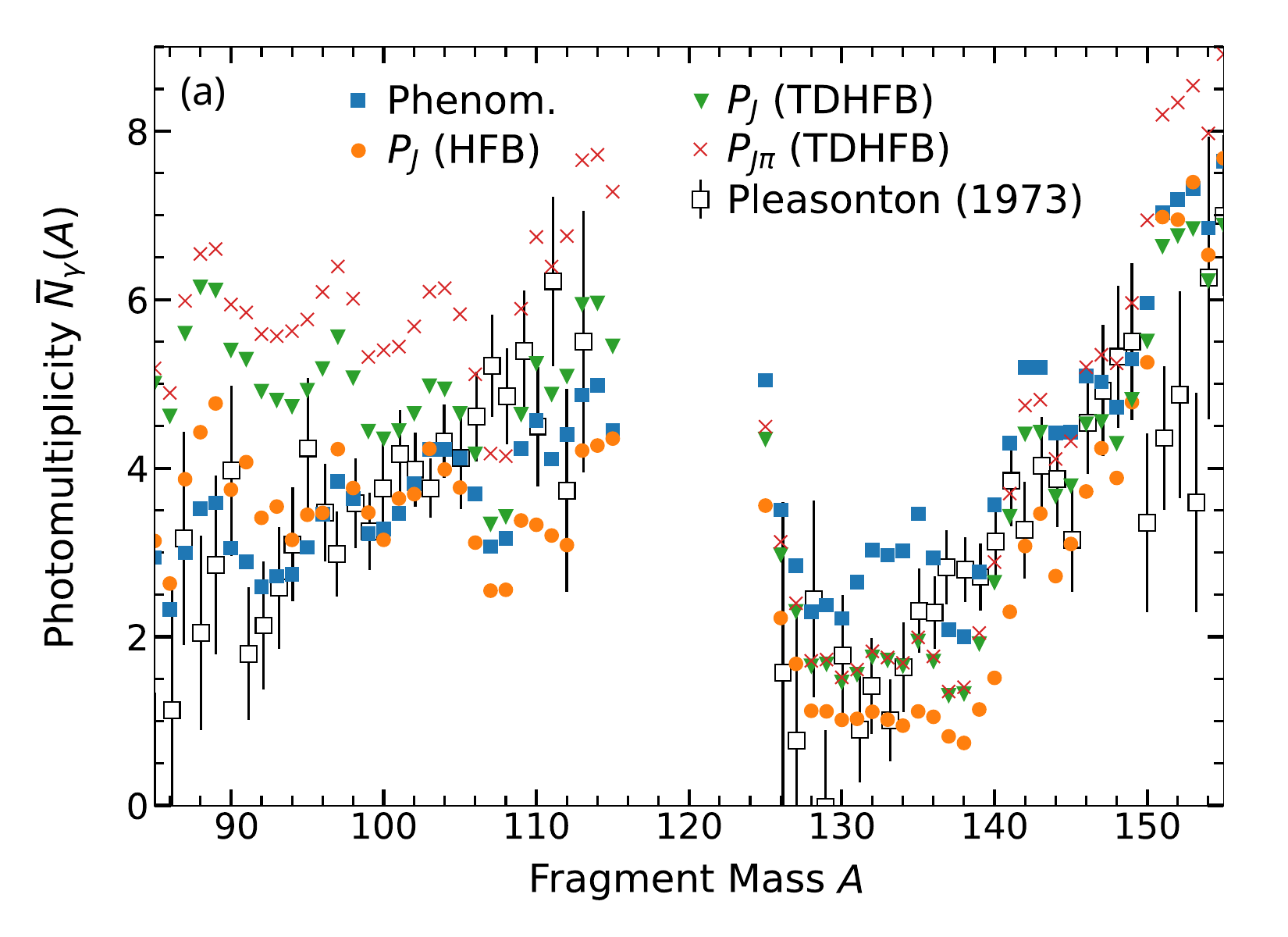}
    \includegraphics[width=0.97\linewidth]{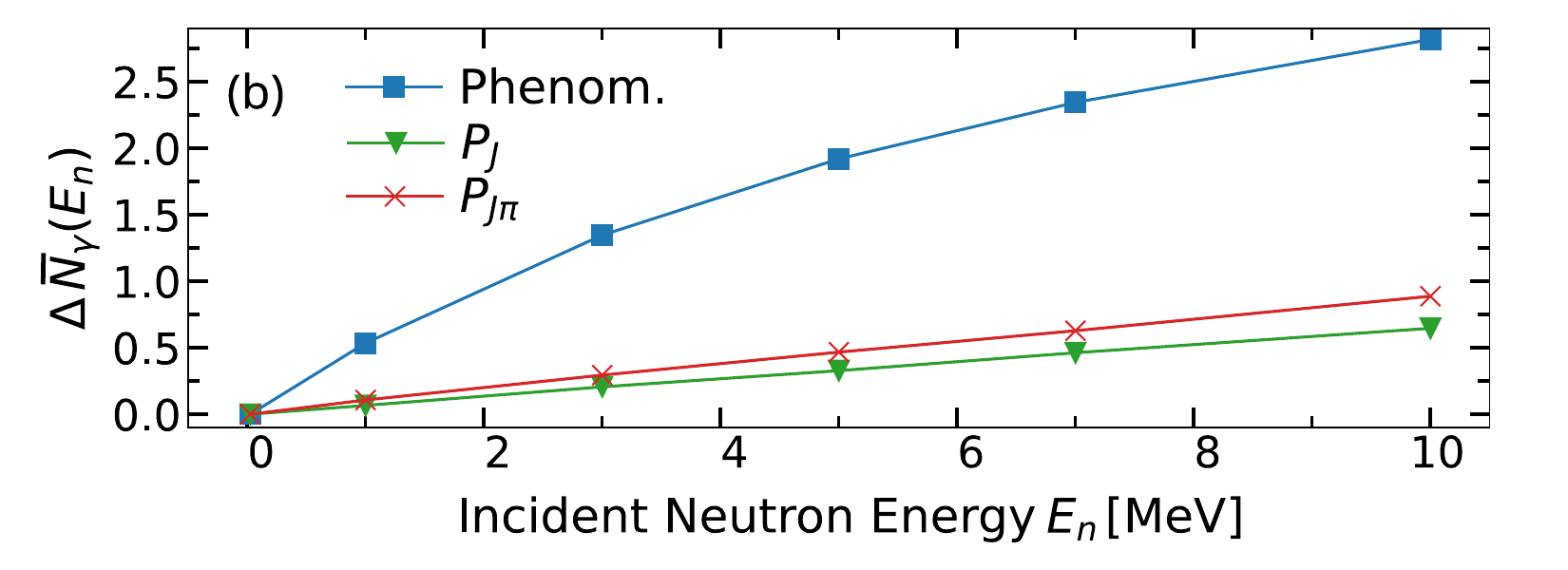}
    \caption{(a) Calculated average photon multiplicity per fragment, $\bar{N}_\gamma(A)$, 
    for thermal fission of $^{239}$Pu, obtained with the \textsc{feta} 
    framework and four prescriptions described in the text.
    Experimental data are taken from Ref.~\cite{pleasanton1973}.
    (b) Total photon multiplicity, relative to the thermal value,
    as a function of incident neutron energy for the phenomenological
    and TDHFB distributions. }
    \label{fig:multiplicities} 
\end{figure}

Figure~\ref{fig:multiplicities}(a) shows the average photon multiplicity 
per fragment mass, $\bar{N}_{\gamma}(A)$, for the four prescriptions 
\cite{SM}. All four reproduce the characteristic sawtooth pattern in 
fragment mass, but differ substantially in magnitude, reaching nearly 
a factor of two in some fragments. This holds true even for the two models 
that yield similar $\bar{N}_{\gamma}$ values, illustrating that reproducing
only the total multiplicity does not guarantee the correct distribution 
across fragment mass. The fact that the two TDHFB predictions differ 
across the mass range is a direct, observable consequence of the spin-parity 
correlation discussed above: even where the marginal parity distribution is 
close to equiprobable, accounting for this correlation can substantially 
modify photon emission.

The evolution of $\bar{N}_\gamma$ with $E_n$ is not well constrained 
\cite{stetcu2020evaluation}. Recently, $\gamma$-multiplicity measurements, 
with a $0.4 < E_\gamma < 2.2$~MeV window, have established a linear 
increase of $N_\gamma$ with $E_n$, with a slope of $0.085 \pm 0.010$ 
MeV$^{-1}$ below second-chance fission \cite{giha2023}. Some phenomenological 
models reproduce this trend reasonably well \cite{giha2023}, even though the
energy dependence of their spin formula is poorly constrained, due to both 
the scarcity of data and its correlation with other fitted quantities 
\cite{stetcu2014}. Figure~\ref{fig:multiplicities}(b) shows the evolution of 
$\bar{N}_\gamma$ with $E_n$, normalized to the corresponding thermal value. 
The phenomenological model employed here predicts a very steep rise of more 
than $2.5$ photons between thermal and $10$~MeV fission. The two microscopic
models, obtained without any parameter tuning to fission data, give a much 
slower, linear increase
of around $0.09$~MeV$^{-1}$.


The present microscopic framework predicts spin and parity distributions 
of fission fragments as functions of incident neutron energy. The spins 
are largely set by the nuclear elongation at scission, while parities 
remain correlated with spins across the entire energy range considered.
Propagating these distributions through a reaction-theory code 
yields testable predictions for photon observables, helping identify 
phenomenological ingredients that require improvement or recalibration.
Future work should assess the impact of these distributions on additional 
fission observables and improve the microscopic treatment of other 
components of the decay model. Alternative ways of imparting excitation 
energy in the microscopic calculations should also be explored, particularly 
those that can disentangle the influence of spin and excitation energy 
of the fissioning system on the spin of fragments \cite{cannarozzo2025}.

\paragraph*{Acknowledgements ---}
The work of P.~M. was funded by the European Union’s Horizon Europe 
research and innovation program under the Marie Sk\l{}odowska-Curie 
Actions Grant Agreement No.~101149053 and by the Croatian Science 
Foundation under the project Dynamics of complex femtosystems 
(IP-2025-02-4214). Support for this work was partly provided through 
Scientific Discovery through Advanced Computing (SciDAC) program 
funded by U.S. Department of Energy, Office of Science, Advanced 
Scientific Computing Research and Nuclear Physics. This work was 
partly performed under the auspices of the US Department of Energy 
by the Lawrence Livermore National Laboratory under Contract 
DE-AC52-07NA27344. Computing support for this work came from the 
Lawrence Livermore National Laboratory Institutional Computing 
Grand Challenge program. G.S. acknowledges support from the French 
National Research Agency (ANR) under the Chaire de Professeur Junior 
programme (Grant No. ANR-23-CPJ1-0154).

\paragraph*{Data Availability ---}
The joint fragment distributions reported in this article will be made 
publicly available upon publication.

\nocite{ring2004,schunck2014,marevic2022,maruhn2014,goddard2015,
kopecky1990test,koning2003local,capote2009ripl}

\bibliography{bibliography}

\ifarXiv
  \clearpage
  \ifarXiv
  \clearpage
  

\section*{Supplementary material}

# Supplemental Material: Energy-Resolved Spin and Parity Distributions of Fission Fragments

Petar Marević
*Physics Department, Faculty of Science, University of Zagreb, 10000 Zagreb, Croatia*

Nicolas Schunck
*Nuclear data and Theory Group, Nuclear and Chemical Science Division, Lawrence Livermore National Laboratory, California, USA 94550*

Antonio Bjelčić
*Applied Mathematics and Computational Research Division, Lawrence Berkeley National Laboratory, California, USA 94720*

Guillaume Scamps
*Université de Toulouse, CNRS/IN2P3, L2IT, Toulouse, France*


This Supplemental Material provides additional details pertinent to the main text, including the choice of basis, the TDHFB propagation, the boost procedure, the projections, the TDGCM weighting, the statistical analysis of the correlations between the scission geometry, the fragment deformation and the fragment spin, and the modeling of the statistical decay.

## I. MEAN-FIELD CALCULATIONS

The theoretical framework is based on Hartree-Fock-Bogoliubov theory (HFB) and its time-dependent extension (TDHFB) [1–3]. Calculations are performed using the AXIALHOHFB computer program [4]. We employ the popular Skyrme SkM* [5] energy density functional (EDF) and neglect the center-of-mass correction term. A mixed volume-surface contact force is used [6], with the typical cutoff in the quasiparticle space of $E_{\text{cut}} = 60$ MeV and the pairing strength parametrization proposed in [7]. Axial symmetry is assumed throughout all calculations.

Single-nucleon wave functions are expanded in the basis of the axially-symmetric harmonic oscillator. This basis is uniquely defined by three parameters: $Z_{\text{box}}, R_{\text{box}}$, and $\max\{n_z\}$. Here, $2Z_{\text{box}}$ is the height of the simulation box cylinder, $2R_{\text{box}}$ is its diameter, and $\max\{n_z\}$ is the largest value of the $n_z$ quantum number present in the basis. More details on the definition of the basis are available in Ref. [4].

In this work, we set $2Z_{\text{box}} = 60$ fm, $2R_{\text{box}} = 25$ fm and $\max\{n_z\} = 38$, both for static and dynamic calculations. This amounts to a total of 5967 basis states. Such a basis provides a very good convergence profile and accurately describes the very elongated post-fission shapes that can emerge on some trajectories. For comparison, a typical basis in all our previous fission studies based on static potential energy surfaces (PES) was comprised of $1100 - 1200$ states; see, e.g., Refs. [7, 8]. Furthermore, state-of-the-art Gogny-TDHFB calculations include 5720 basis states, with a computational cost of about $10^4$ CPU hours per trajectory [9]. Even with $N_{\text{basis}} = 5967$ basis states, the computational cost of TDHFB simulations with AXIALHOHFB is on the order of 100 CPU hours per trajectory. This enables, for the first time, an application of projection techniques for a large number of TDHFB trajectories.

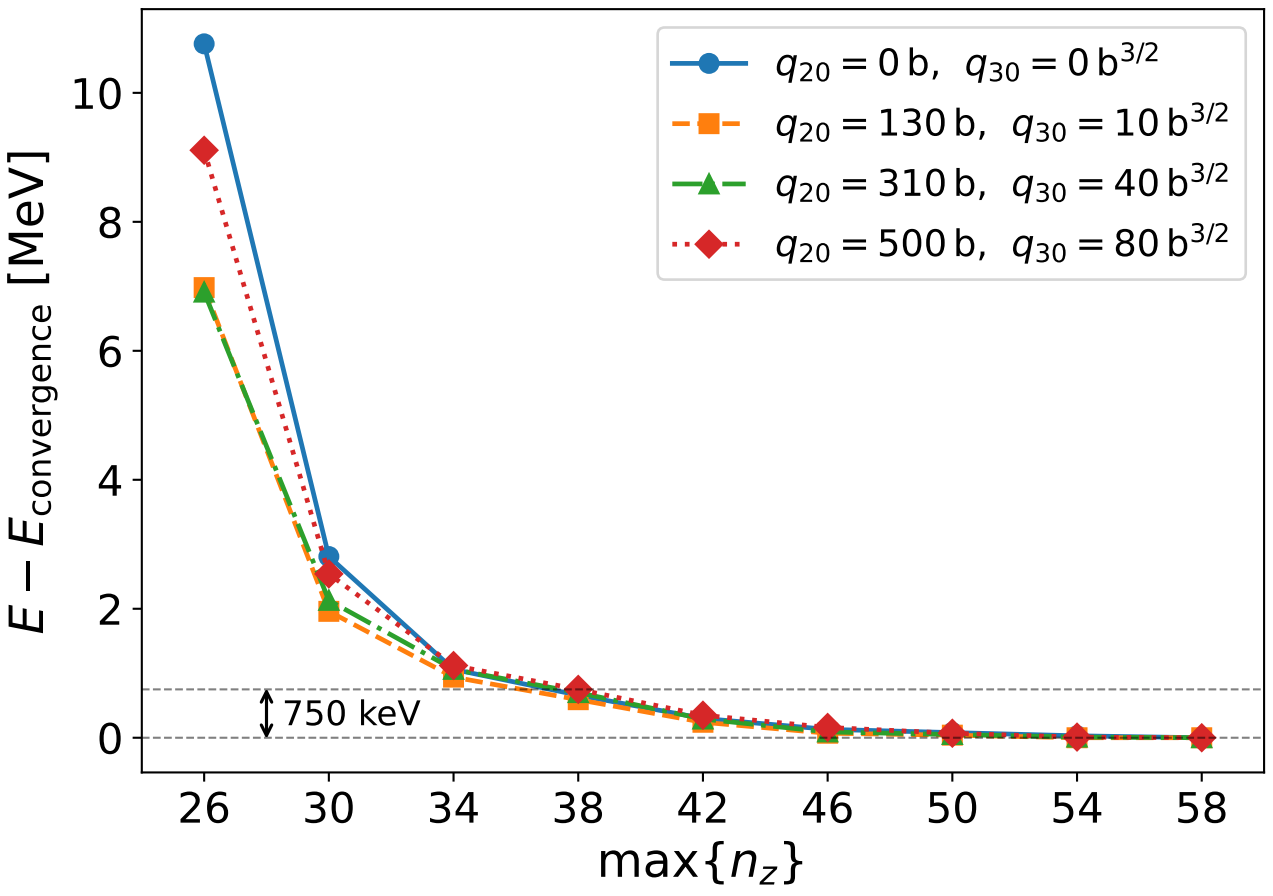


FIG. 1. Energy of four representative HFB configurations as a function of $\max\{n_z\}$, for fixed $2Z_{\text{box}} = 60$ fm and $2R_{\text{box}} = 25$ fm. Energies are shown relative to the corresponding fully converged values, obtained for $\max\{n_z\} = 58$. The value $\max\{n_z\} = 38$ is used in all subsequent calculations. See text for more details.

### A. Static calculations

To demonstrate the convergence of static calculations, we fix the values of two basis parameters ($2Z_{\text{box}} = 60$ fm and $2R_{\text{box}} = 25$ fm) and calculate the HFB binding energy as a function of the third parameter, $\max\{n_z\}$. Constraints are imposed on the expectation values of axially symmetric quadrupole ($q_{20}$) and octupole ($q_{30}$) mo-

ments, which describe, respectively, the elongation and the reflection asymmetry of the nuclear shape. We consider four representative configurations from Ref. [4]: the spherical configuration ($q_{20} = 0$ b, $q_{30} = 0$ b$^{3/2}$), a configuration just beyond the saddle point ($q_{20} = 130$ b, $q_{30} = 10$ b$^{3/2}$), a near-scission configuration ($q_{20} = 310$ b, $q_{30} = 40$ b$^{3/2}$), and a fully fragmented configuration ($q_{20} = 500$ b, $q_{30} = 80$ b$^{3/2}$). In Fig. 1 we show the HFB energies of these configurations calculated with $\max\{n_z\} \in [26, 58]$, relative to the fully converged values obtained with $\max\{n_z\} = 58$. All four configurations exhibit a very similar convergence profile. The $\max\{n_z\} = 58$ calculations are converged at the sub-1 keV level. However, the corresponding basis is comprised of $N_{\rm basis} = 20709$ states, which is well beyond current computational capabilities. An optimal compromise between numerical accuracy and computational cost is achieved for $\max\{n_z\} = 38$, which provides a sub-750 keV convergence for all four configurations, amounting to less than 0.05% of the total binding energy. Note that the same basis is used for all static configurations and throughout dynamic simulations. While it differs from the common practice of adjusting basis parameters at each constrained deformation, our convergence analysis indicates that these numerical results are, in fact, likely better converged than most previous studies published in the literature [4].

### B. Dynamic calculations

#### 1. *Baseline trajectories*

TDHFB evolution is simulated using the AXIALHOHFB program [4]. First, we calculate the HFB potential energy surface (PES) in the $(q_{20}, q_{30})$ plane, as shown in Fig. 2. Since TDHFB does not allow for tunneling through barriers, initial points for TDHFB simulations are conventionally taken beyond the saddle point. In this work, we choose a set of 10 initial configurations in $^{240}$Pu, marked with circles in Fig. 2. Their energies are in the window of $[6.35, 6.65]$ MeV above the ground state, simulating the thermal fission of $^{239}$Pu. Their quadrupole deformation $q_{20}$ and octupole deformation $q_{30}$ are summarized in Table I.

The initial configurations are evolved in time according to the TDHFB equations using a fourth-order Runge-Kutta method with time step $\Delta t = 0.5$ fm/c. The resulting fission trajectories, traced by the $(q_{20}, q_{30})$ values of the nucleus, are superimposed on the PES in Fig. 2. Note that even the trajectories that start at relatively low $q_{30}$ values do not end up in symmetric fission, which is consistent with a negligible symmetric contribution at thermal energies. The time evolution of each trajectory is simulated up to the point when the density of either primary fragment reaches the edge of the box. In our calculation, this is defined as the point when the on-axis isoscalar density $\rho(z, r_\perp = 0)$ exceeds 0.01 fm$^{-3}$ within $\Delta z = 4$ fm of either edge. The corresponding distance $d$ between the centers of mass of both FFs at that point, as well as the the average neutron and proton numbers in FFs, are also summarized in Table I.

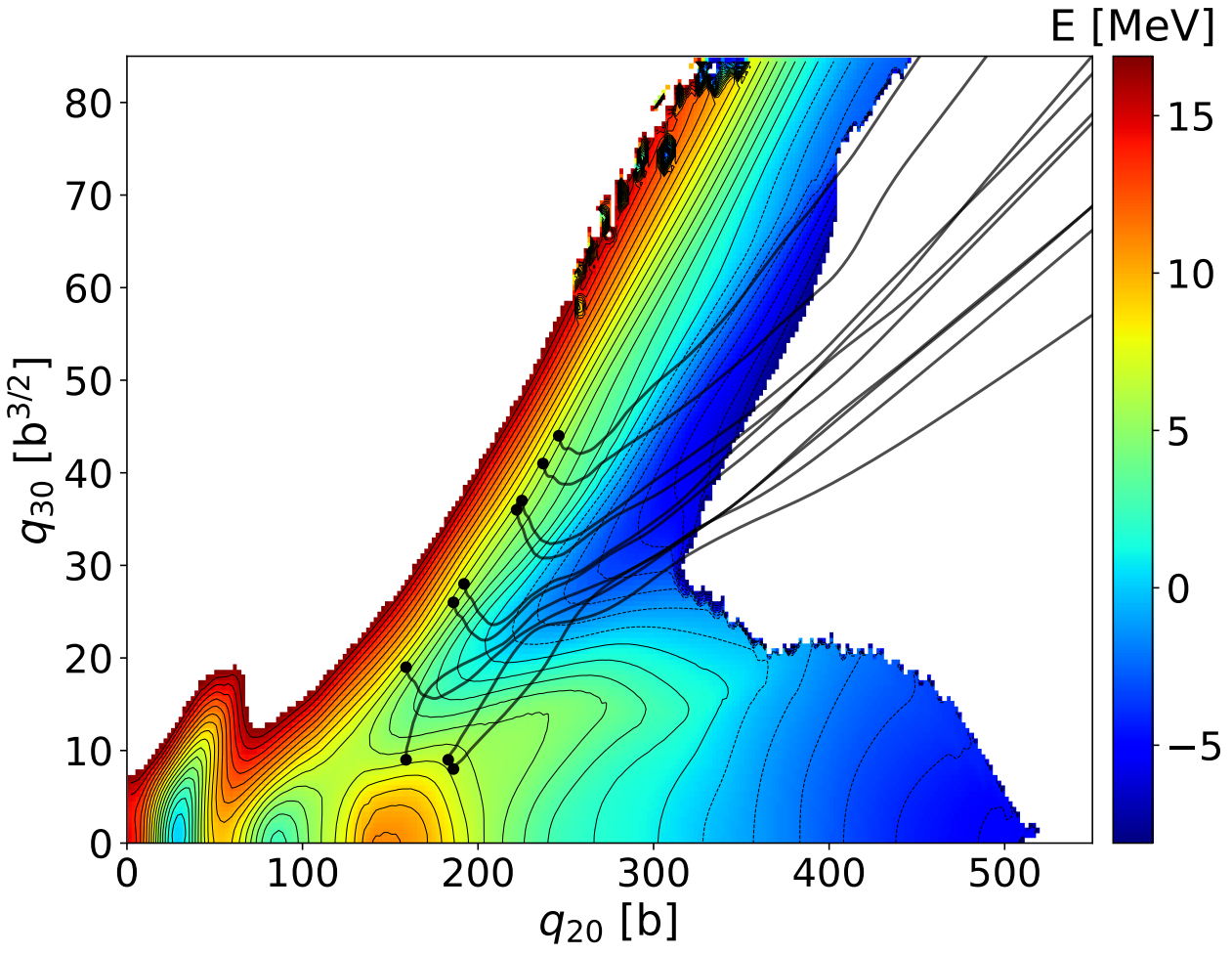


FIG. 2. Potential energy surface of $^{240}$Pu calculated with the Skyrme SkM* EDF, overlayed with 10 TDHFB trajectories, corresponding to thermal fission. See text for more details.

TABLE I. Properties of 10 baseline TDHFB trajectories. The quadrupole moment is in units of barns (b), and the octupole moment in units of b$^{3/2}$. Average numbers of particles are dimensionless.

| Traj. | $(q_{20}, q_{30})$ | $\langle N_H\rangle, \langle Z_H\rangle$ | $\langle N_L\rangle, \langle Z_L\rangle$ | $d$ [fm] |
|---|---|---|---|---|
| 1 | (159.0, 9.0) | 83.2, 52.6 | 62.8, 41.4 | 31.0 |
| 2 | (159.0, 19.0) | 84.3, 53.1 | 61.7, 40.9 | 31.4 |
| 3 | (183.0, 9.0) | 80.9, 51.6 | 65.1, 42.4 | 31.0 |
| 4 | (186.0, 8.0) | 81.9, 52.4 | 64.1, 41.6 | 30.0 |
| 5 | (186.0, 26.0) | 85.6, 54.0 | 60.4, 40.0 | 30.1 |
| 6 | (192.0, 28.0) | 86.9, 54.9 | 59.1, 39.1 | 29.5 |
| 7 | (222.0, 36.0) | 84.3, 53.4 | 61.7, 40.6 | 29.7 |
| 8 | (225.0, 37.0) | 85.5, 53.8 | 60.5, 40.2 | 29.5 |
| 9 | (237.0, 41.0) | 90.3, 56.8 | 55.7, 37.2 | 28.2 |
| 10 | (246.0, 44.0) | 93.9, 58.6 | 52.1, 35.4 | 27.8 |

The total energy of the nucleus is conserved in TDHFB theory. However, the inclusion of a quasiparticle cutoff, which is a common way to avoid divergences in contact pairing interactions, renders the full Bogoliubov matrix formally non-unitary. Consequently, the total energy is not exactly conserved. In practice, for large bases and cutoffs such as the ones employed here, this violation is very modest [4]. In Fig. 3, we show the energy variation $E(t) - E(0)$ for all trajectories. Overall, the violation is rather small and it always stays below 150 keV, which amounts to less than 0.01% of the total binding energy.

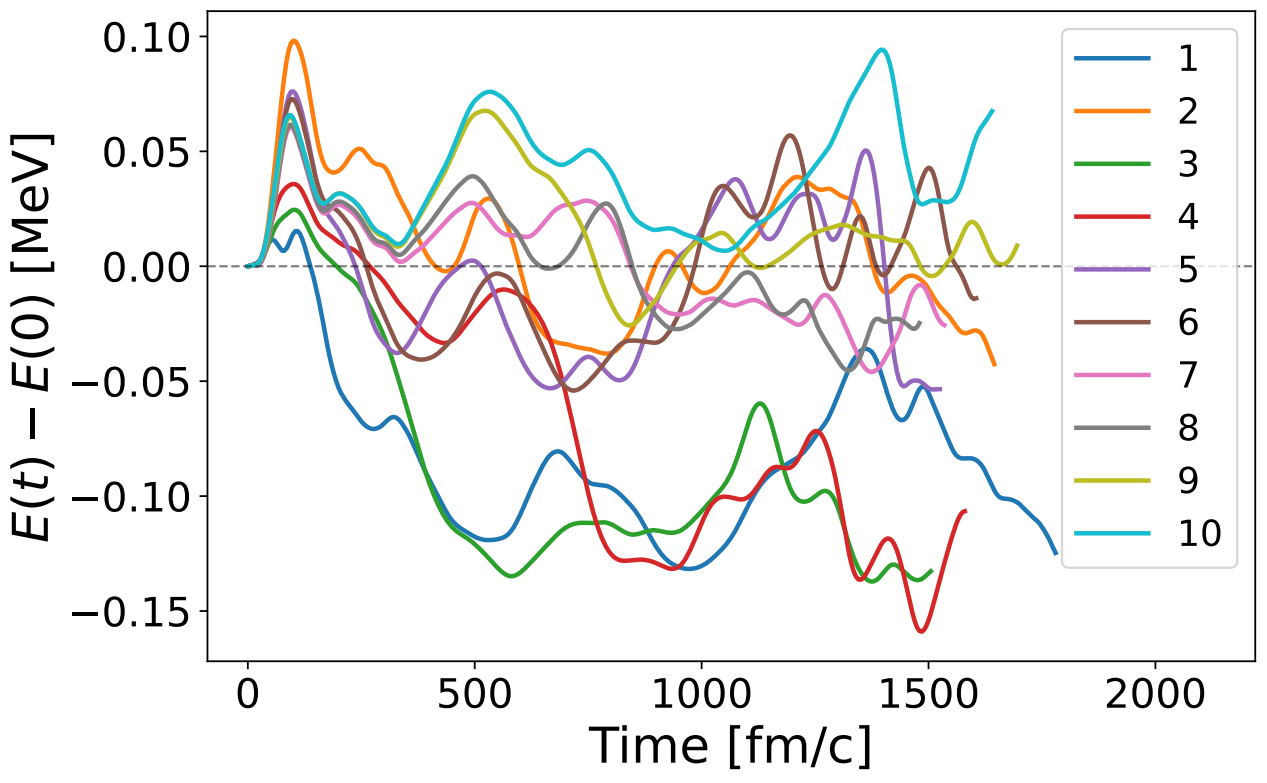


FIG. 3. Energy conservation along TDHFB trajectories from Fig. 2. See text for more details.

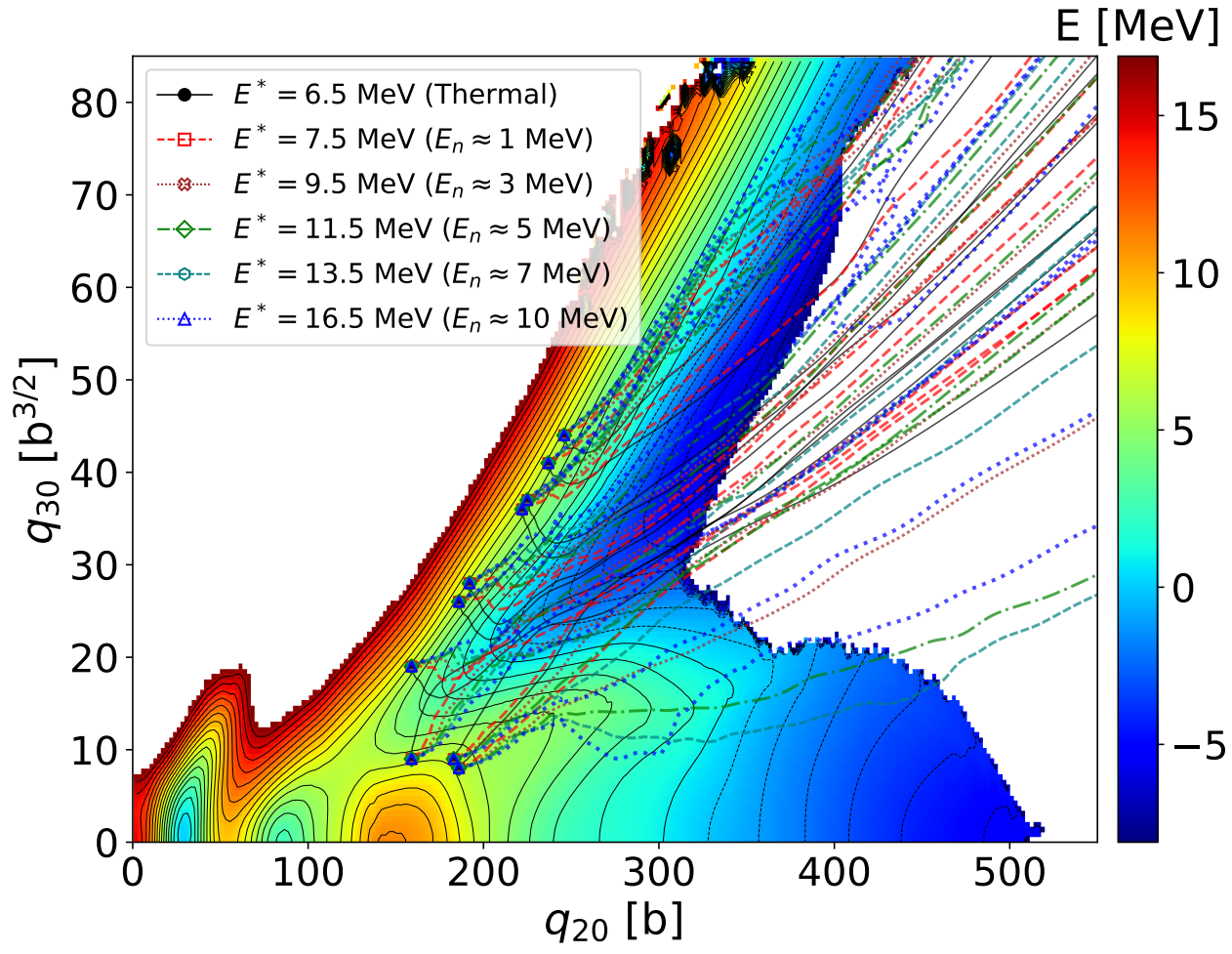


FIG. 4. Potential energy surface of $^{240}$Pu calculated with the Skyrme SkM* EDF, overlayed with 60 TDHFB trajectories, corresponding to thermal and $E_n = 1, 3, 5, 7, 10$ MeV fission. See text for more details.

### 2. *Excited trajectories*

The effects of intrinsic nuclear excitation are simulated by applying the instantaneous quadrupole boost to the initial configurations [2, 10]. By choosing the appropriate value of boost parameter, an HFB configuration can be boosted to the excitation energy $E^*$ above the ground state. The equivalent energy of the incident neutron can then be estimated as $E_n = E^* - S_n$, where $S_n$ is the one-neutron separation energy, equal to roughly 6.5 MeV in this case.

In a series of calculations, the set of initial configurations was boosted to the excitation energies corresponding to $E_n = 1, 3, 5, 7, 10$ MeV. The resulting configurations are then propagated in time using the same method and parameters as in the baseline calculation. In Fig. 4 we show all the trajectories. With increasing $E_n$ some trajectories shift toward lower $q_{30}$ values, corresponding to the increasing population of symmetric fission.

To study the effect of instantaneous quadrupole boost on the nuclear wave function, in Fig. 5 we show $\Delta E(t) = E_{\text{boost}}(t) - E_{\text{baseline}}(t)$, the difference in various energy components between the boosted ($E_n = 10$ MeV) trajectory and the corresponding baseline trajectory, for all initial configurations. The following energy components are shown separately:

- The time-even contribution $E_{\text{even}} = E_{\rho\rho} + E_{\rho\Delta\rho} + E_{\rho\tau} + E_{\rho\nabla J}$,
- The time-odd contribution $E_{\text{odd}} = E_{ss} + E_{s\Delta s} + E_{sT} + E_{s\times j}$,
- The current density contribution $E_{jj}$, which is also time-odd,
- The Coulomb contribution $E_{\text{Coul}}$,
- The pairing contribution $E_{\text{pair}}$,
- The total kinetic energy contribution $E_{\text{kin}}$,
- The collective kinetic energy contribution $E_{\text{coll.kin.}}$,
- The total energy $E_{\text{total}}$.

Various components of an EDF are defined, e.g., in [11]. The collective kinetic energy can be calculated as

$$E_{\text{coll.kin.}} = \frac{\hbar^2}{2m} \int d^3\boldsymbol{r} \frac{\boldsymbol{j}^2(\boldsymbol{r})}{\rho(\boldsymbol{r})}, \tag{1}$$

where $\boldsymbol{j}(\boldsymbol{r})$ is the current density and $\rho(\boldsymbol{r})$ is the particle density [12, 13]. In contrast to the total kinetic energy, which includes contributions of both internal excitations and collective motion, $E_{\text{coll.kin.}}$ contains only the contribution from the latter.

All trajectories in Fig. 5 exhibit a remarkably similar pattern. At $t = 0$, the instantaneous quadrupole boost induces internal currents, which are reflected in non-zero values of $E_{jj}$ and $E_{\text{kin.coll}}$ for boosted configurations. On the other hand, the EDF components that depend only on the particle, spin, or pairing density are unchanged at $t = 0$. However, the energy provided by the boost is quickly transferred into the nuclear terms of an EDF. The $\Delta E_{\text{even}}$ rises rapidly, reaching its maximum at about $t = 60$ fm/c. It then drops sharply and continues oscillating with a modest amplitude. Variations in $\Delta E_{\text{odd}}$ are generally rather small. Since the prefragments are pushed away by the initial boost, the Coulomb energy first drops sharply. It then rises again under the influence of the attractive nuclear force, undergoes several oscillations, and eventually increases for all trajectories. The different asymptotic values reflect the different regions of the PES in which the trajectories end up. The total energy difference, shown in the last panel, remains constant at the value of the imparted excitation energy throughout the evolution, confirming that the boost injects the intended amount of energy and that it is conserved thereafter.

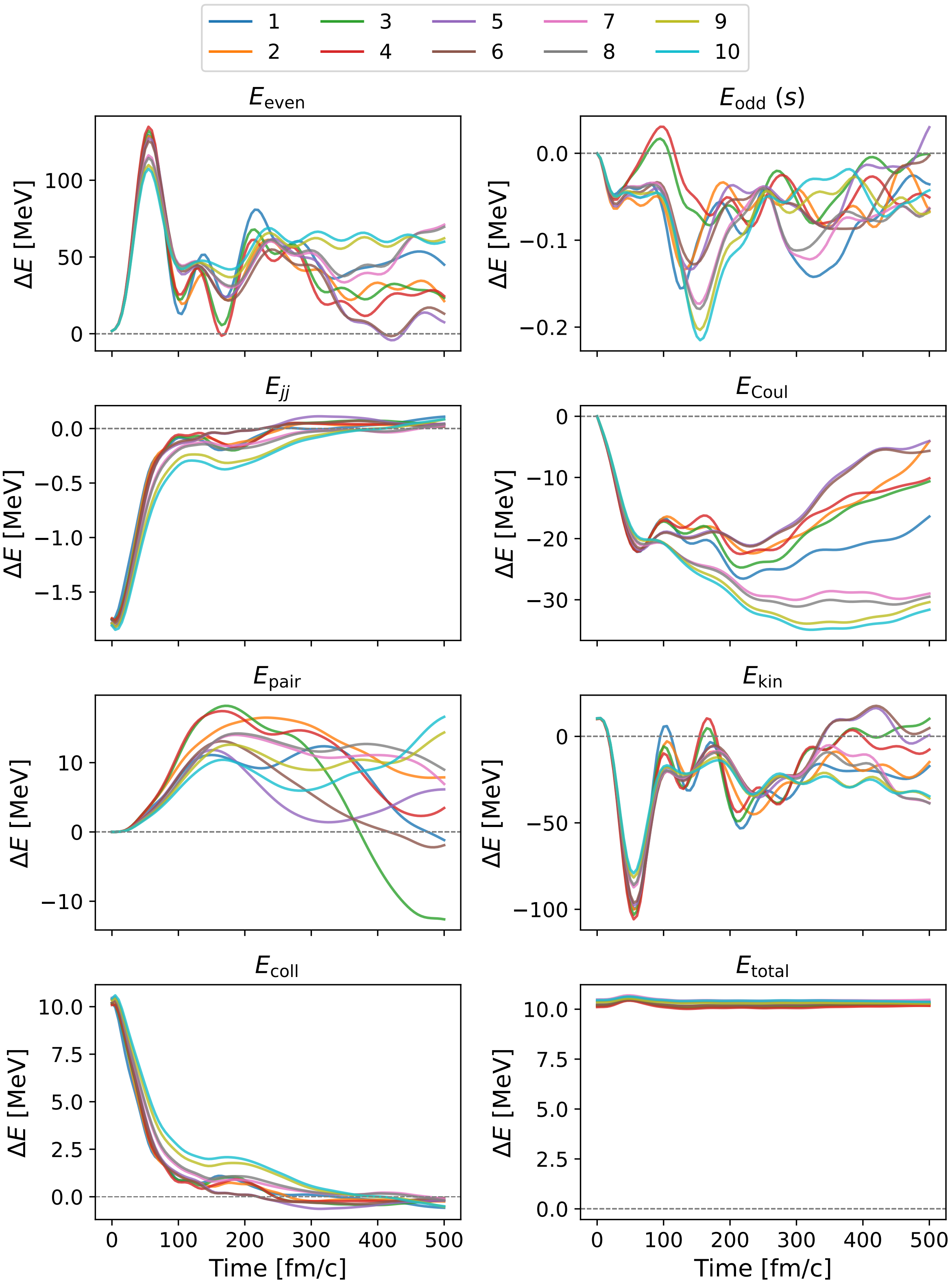


FIG. 5. Difference in various energy components between the boosted ($E_n = 10$ MeV) and the corresponding baseline trajectory, $\Delta E(t) = E_{10\ \mathrm{MeV}}(t) - E_{\mathrm{thermal}}(t)$, for all ten initial configurations. The components are defined in the text. Note that the total energy difference remains constant at the value of the imparted excitation energy.

Overall, this analysis is in agreement with conclusions of Ref. [10]. The initial excitation energy is first deposited as the collective kinetic energy, but is very quickly ($t < 1$ zs) transferred into the nuclear terms of the EDF. This reflects the rapid dispersal of the initially coherent nucleon current $\boldsymbol{j}(\mathbf{r})$ into incoherent single-particle motion, after which the system retains little memory of the excitation mechanism. True thermalization through two-body collisions, absent in TDHFB, would only accelerate this process. Combined with the fact that the nucleus must undergo a large increase of $q_{20}$ on the way to scission, this justifies using quadrupole boosts to model nuclear fission at different excitation energies.

## II. PROJECTIONS TECHNIQUES AND TIME-DEPENDENT GENERATOR COORDINATE METHOD

The full-fledged distribution in fission fragment angular momentum $J_F$, its component along the fission axis $K_F$, parity $\pi_F$, neutron and proton numbers, $N_F$ and $Z_F$, as well as the neutron and proton number of the total fissioning system, $N$ and $Z$, is obtained as

$$\mathbb{P}_F(J_F, K_F, \pi_F, N_F, Z_F, N, Z|q) = \langle \Phi_q | \hat{P}^{J_F}_{K_F K_F} \hat{P}^{\pi_F} \hat{P}^{N_F} \hat{P}^{Z_F} \hat{P}^{N} \hat{P}^{Z} | \Phi_q \rangle \,, \tag{2}$$

where $F = l, r$ stands for the left or the right fragment, $|\Phi_q\rangle$ is the TDHFB state on trajectory $q$, $\hat{P}^{J_F}_{K_F K_F}$ is the angular momentum projection (AMP) operator [14], $\hat{P}^{\pi_F}$ is the parity projection operator [15], while $\hat{P}^{N_F, Z_F}$ and $\hat{P}^{N,Z}$ are the particle number projection (PNP) operators numbers in the fragments and the total system, respectively [16].

In AMP, an equidistant mesh of $N_\alpha = N_\gamma = 18$ points is used for Euler angles $\alpha \in [0, 2\pi]$ and $\gamma \in [0, 4\pi]$, while the Gauss-Legendre integration with $N_\beta = 42$ points is used for the Euler angle $\beta \in [0, \pi]$. PNP is in all cases performed with $N_\varphi = 17$ equidistant points in the gauge angle $\varphi \in [0, 2\pi]$.

In practice, we keep only the component of the distribution (2) corresponding to the correct total number of particles, i.e. $(N, Z) = (146, 94)$ for $^{240}$Pu$^*$. This gives us the distribution $\mathbb{P}_F(J_F, K_F, \pi_F, N_F, Z_F|N, Z, q)$, which is re-normalized to 1. Such distributions are obtained for each trajectory $q$, at each energy $E_n$. In the following, for simplicity, we do not explicitly denote projections on the total number of particles, but they are carried out in all cases. Consequently, quantities $J, K, \pi, N, Z$ all refer to the properties of fission fragments in the remainder of the document. Similarly, we do not denote explicit dependence on $E_n$, but we underline that all the quantities are energy dependent.

In the next step, we use the time-dependent generator coordinate method under the Gaussian overlap approximation [17, 18] to determine the weight $F(q)$ associated at each trajectory, such that $\sum_q F(q) = 1$ at each $E_n$. Summing up over all trajectories we obtain

$$\mathbb{P}_F(J, K, \pi, N, Z) = \sum_q F(q) \mathbb{P}_F(J, K, \pi, N, Z|q). \tag{3}$$

By adding up the contributions from both fragments we get

$$\begin{aligned} \mathbb{P}(J, K, \pi, N, Z) = &\, \mathbb{P}_l(J, K, \pi, N, Z) \\ &+ \mathbb{P}_r(J, K, \pi, N, Z). \end{aligned} \tag{4}$$

The distribution (4) is, by construction, normalized to 2. The final distribution is obtained by renormalizing it to 1. In this final distribution, we keep only the fragments $(N_0, Z_0)$ whose relative weight

$$W(N_0, Z_0) = \sum_{J,K,\pi} \mathbb{P}(J, K, \pi | N_0, Z_0) \tag{5}$$

exceeds some small $\epsilon$. In practice, we set $\epsilon = 10^{-6}$.

## III. CORRELATION BETWEEN SCISSION GEOMETRY, FRAGMENT DEFORMATION AND FRAGMENT SPIN

### A. Definitions

In this section we provide further details on the correlation between scission geometry, fragment deformation and fragment spin. The mean spin of the fragment $(N, Z)$ is calculated as

$$\langle J \rangle_{N,Z} = \sum_J J \, \mathbb{P}(J|N, Z), \tag{6}$$

where $\mathbb{P}(J|N, Z) = \sum_{K,\pi} \mathbb{P}(J, K, \pi|N, Z)$ is the corresponding spin distribution, normalized to 1, at fixed $E_n$. The yield-weighted average is then obtained as

$$\langle J \rangle = \sum_{N,Z} \langle J \rangle_{N,Z} \, \mathbb{P}(N, Z), \tag{7}$$

where $\mathbb{P}(N, Z) = \sum_{J,K,\pi} \mathbb{P}(J, K, \pi, N, Z)$ is the calculated yield, i.e. the probability that the fragment $(N, Z)$ is formed, normalized to 1.

We consider ten trajectories $q$ at each energy $E_n$. Each of them can be characterized by the quadrupole and octupole deformation of the fissioning system at scission, $q_{20}(q)$ and $q_{30}(q)$, as well as the TDGCM weight $w(q)$.

We can therefore estimate the average deformations at scission as

$$\langle q_{20}\rangle = \sum_q q_{20}(q)w(q), \tag{8}$$

$$\langle q_{30}\rangle = \sum_q q_{30}(q)w(q). \tag{9}$$

Finally, the fragment average quadrupole deformation parameter $\langle\beta_2\rangle$ is defined as

$$\langle\beta_2\rangle(N,Z) = \frac{\sum_q w(q)\beta_2(N,Z,q)}{\sum_q w(q)s(N,Z,q)}, \tag{10a}$$

$$\beta_2(N,Z,q) = \mathbb{P}_l(N,Z|q)\beta_2^l(q) + \mathbb{P}_r(N,Z|q)\beta_2^r(q), \tag{10b}$$

$$s(N,Z,q) = \mathbb{P}_l(N,Z|q) + \mathbb{P}_r(N,Z|q), \tag{10c}$$

where $\beta_2^F(q)$ is the quadrupole deformation of the left $(F=l)$ or the right $(F=r)$ fragment on trajectory $q$ and

$$\mathbb{P}_F(N,Z|q) = \sum_{J,K,\pi} \mathbb{P}_F(J,K,\pi,N,Z\,|\,q) \tag{11}$$

is the distribution of probability that the fragment $(N,Z)$ is formed on trajectory $q$ in fragment $F$.

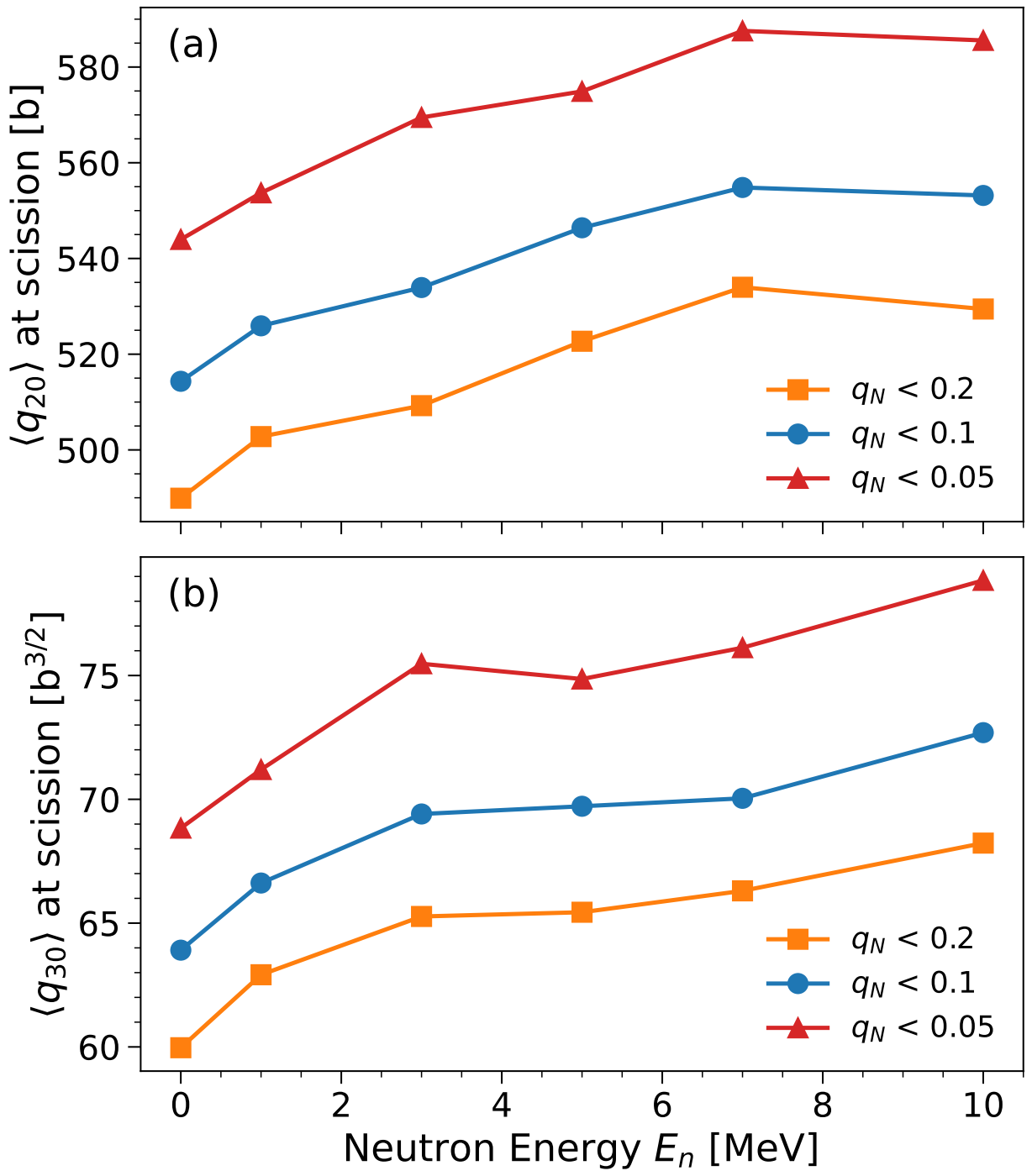


FIG. 6. Average quadrupole (a) and octupole (b) deformation of the fissioning system at scission, as a function of incident neutron energy $E_n$. Three choices of scission threshold are considered, $q_N = 0.2, 0.1, 0.05$, where $q_N$ is the neck value.

### B. Scission elongation and the mean spin

In Fig. 6 we show the average quadrupole and octupole deformation of the fissioning system at scission, as a function of incident neutron energy $E_n$. Here, scission is defined as the point where the expectation value of the neck operator $q_N$ drops below a certain threshold. The results are robust with respect to the choice of this threshold and show the same qualitative behavior for $q_N = 0.2, 0.1, 0.05$. In particular, we find that the quadrupole deformation increases up to $E_n \approx 7$ MeV, after which it plateaus or even decreases. The octupole deformation, on the other hand, continues increasing.

The evolution of the quadrupole deformation at scission, corresponding to the elongation of the nuclear shape, closely follows that of the yield-weighted mean spin, shown in Fig. 2(e) of the main text. This correspondence is quantified in Fig. 7, where we fit $\langle J\rangle$ linearly against $E_n$ and against $\langle q_{20}\rangle$ separately. The latter gives a markedly better fit ($R^2 = 0.993$ vs $R^2 = 0.860$), indicating that $\langle q_{20}\rangle$ is the more proximate predictor of the mean spin. This is consistent with a picture in which the neck ruptures at a configuration whose elongation becomes largely independent of $E_n$ above $\approx 7$ MeV, so that fragment properties tied to that configuration – deformation, and through it, spin – saturate together.

We further tested this link at the level of individual trajectories: at each fixed $E_n$, trajectories that reach scission at larger $q_{20}$ produce higher-spin fragments than trajectories scissioning at smaller $q_{20}$, with a Pearson correlation coefficient $r$ in the range 0.86–0.94 across all six energies. Since this correlation holds for a fixed $E_n$, it cannot be explained by both quantities simply growing with $E_n$ independently, and instead supports a direct link between scission shape elongation and fragment spin.

### C. Fragment deformation and fragment spin

Next, we quantify the link between fragment deformation and spin. For each incident neutron energy $E_n$, we calculate $\Delta\langle\beta_2\rangle(N,Z) = \langle\beta_2\rangle_{E_n} - \langle\beta_2\rangle_{\text{Thermal}}$ and $\Delta\langle J\rangle(N,Z) = \langle J\rangle_{E_n} - \langle J\rangle_{\text{Thermal}}$ for every fragment $(N,Z)$ populated at both energies. The Pearson correlation coefficient of $\Delta\langle\beta_2\rangle$ against $\Delta\langle J\rangle$ is then computed over more than 400 fragments available at each energy. This yields $r = 0.74$–$0.90$, an indicator of a strong positive correlation.

As a cross-check, we repeat the analysis using the per-trajectory values of $\beta_2^F(q)$ and $\langle J\rangle^F(q) = \sum_J Jp(J|q)$ of fully separated TDHFB fragments, which are not smoothed by TDGCM weighting. $\Delta\beta_2^F(q)$ and $\Delta\langle J\rangle^F(q)$ are calculated for each of the 10 trajectories, at each of the 5 excitation energies, for each of the 2 fission fragments (100 points total), giving a pooled Pearson correlation $r = 0.83$ (cluster bootstrap over trajectories, $10^4$ resamples: 95% confidence interval $[0.63, 0.92]$).

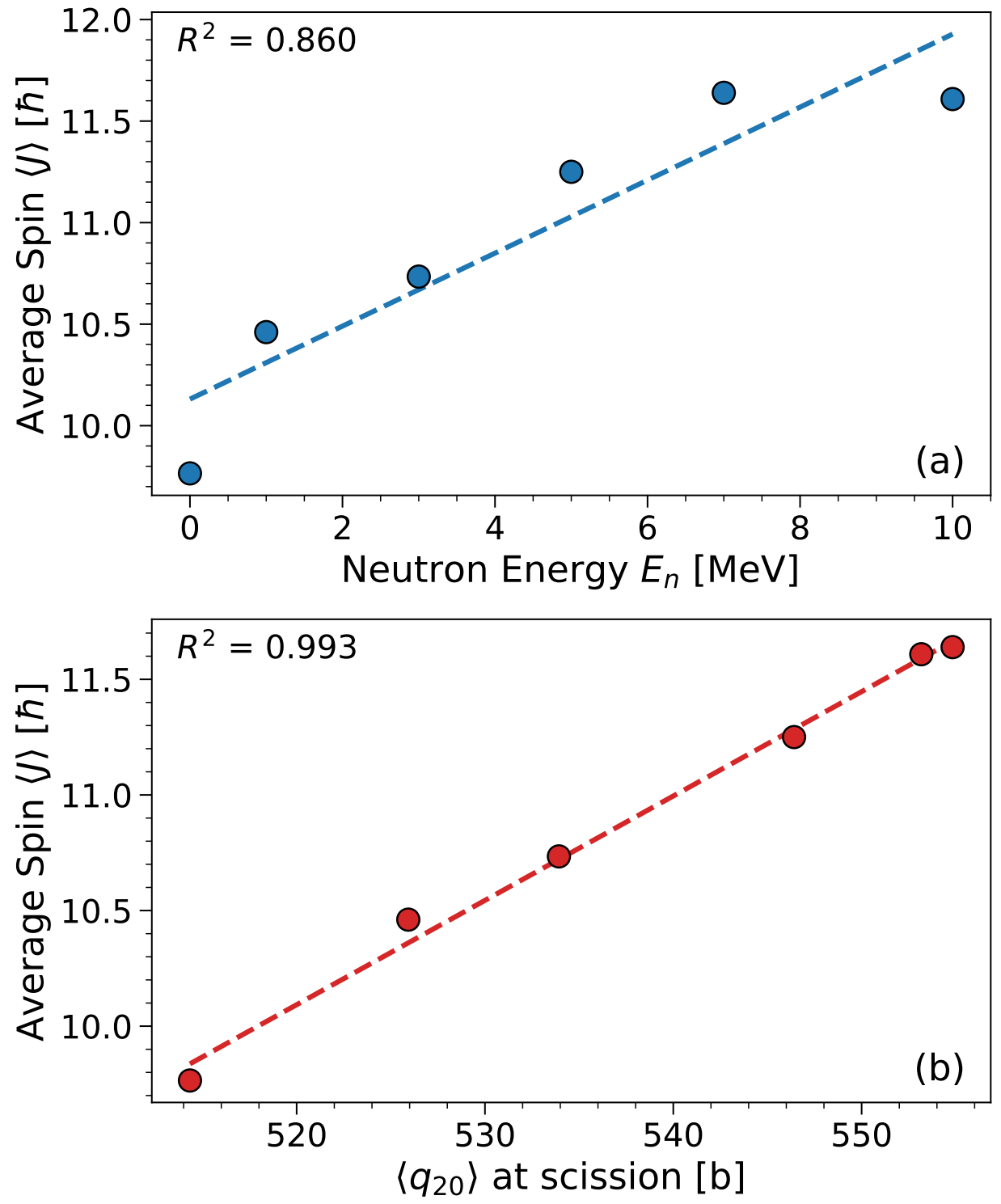


FIG. 7. Average spin $\langle J \rangle$ as a function of incident neutron energy $E_n$ (a) and of the average quadrupole deformation at scission $\langle q_{20} \rangle$ (b), together with a linear fit to each (dashed lines). The scission quadrupole deformation is a better predictor of $\langle J \rangle$ than the incident energy ($R^2 = 0.993$ vs. $R^2 = 0.860$).

Since both quantities grow with $E_n$ for most fragments, this pooled correlation alone does not establish a genuine deformation-spin link, as it could simply reflect a shared trend with $E_n$. We rule this out in two independent ways. First, computing the correlation within each fixed $E_n$ separately (twenty points per energy, where $E_n$ cannot inflate the result) gives $r = 0.76$–$0.88$. Second, computing the partial correlation between the residuals of $\Delta\beta_2^F(q)$ and $\Delta\langle J \rangle^F(q)$ after regressing each against $E_n$ gives $r = 0.82$. Both remain close to the pooled value of 0.83, confirming that the deformation-spin correlation is genuine and not merely an artifact of both quantities increasing with excitation energy.

## IV. FISSION FRAGMENT DEEXCITATION

The deexcitation of fission fragments is modeled with the Hauser-Feshbach theory of statistical nuclear reactions. We use the YAHFC (Yet Another Feshbach Code) [19] code to deexcite each fission fragment and the FETA [20] (Fission Evaluation Tools and Analytics) to define the initial conditions for all fission fragments and reconstruct fission observables based on the decay of all the fragments. In all calculations, the continuum part of the energy spectrum was discretized in bins of 200 keV and each deexcitation proceeded through 100,000 Monte Carlo events. Here, an event refers to an entire deexcitation path from an initial state of energy $E$ and spin-parity $J^\pi$ to a final state where no more decay is possible. Therefore, each event typically includes multiple particle emissions. In our simulations, we did not account for pre-scission neutron emission, nor did we consider multi-chance fission: all the decays correspond to the fragments of the $^{240}$Pu$^*$ compound nucleus.

Figure 4 in the main manuscript reports the average number of photons per fragment mass, $\bar{N}_\gamma(A)$. This number is obtained as

$$\bar{N}_\gamma(A) = \frac{\sum_{Z+N=A} Y_{\rm pre}(Z,N)\,\langle\gamma(Z,N)\rangle}{\sum_{Z+N=A} Y_{\rm pre}(Z,N)} \tag{12}$$

where $\langle\gamma(Z,N)\rangle$ is the average number photons emitted by the fragment $(Z,N)$ over all the 100,000 Monte Carlo events, and $Y_{\rm pre}(Z,N)$ the initial yield of that nucleus.

Fission fragment deexcitation requires two classes of inputs. The initial distribution of fission fragments $Y_{\rm pre}(Z,N)$ prior to any emission of particles, the excitation energy $E^*(Z,N)$ of each fragment and the spin distribution $\mathbb{P}(J|Z,N)$ of all fragments are provided by fission models. The focus of this work is the optimal determination of $\mathbb{P}(J|Z,N)$. For the two other inputs, we use the default models for the $^{239}$Pu(n,f) reaction that are implemented in the CGMF code [21]: the function $Y_{\rm pre}(A)$ is parameterized by a sum of five-Gaussians, the parameters of which are themselves function of the energy of the incident neutron, with the Wahl systematics giving $\mathbb{P}(Z|A)$ in order to reconstruct the full 2D distribution. The excitation energy of the fragments is obtained by invoking a sharing mechanism of the total excitation energy between the heavy and light fragment. The total excitation energy of the reaction is itself obtained from energy conservation and a fit to the experimentally measured total kinetic energy (TKE) as a function of the mass of the fragment; see [21] for a description of all these models. The values of the parameters are the ones shipped with CGMF and contained in the files `yamodel.dat` (mass distributions), `tkemodel.dat` (fit of the TKE) and `rta.dat` (sharing mechanism). The spin distributions are taken from the parameterization implemented in the `setSpinStrength()` function of the file `FissionFragments.cpp` of the CGMF source code; see https://github.com/lanl/cgmf/tree/master.

The second class of inputs is provided by nuclear structure models that will determine how the statistical decay proceeds. These include nuclear level densities, $\gamma$-strength functions, transmission coefficients, and low-lying discrete excited states, electromagnetic transitions and branching ratios. Since the purpose of this work is to examine the impact of different spin-parity distributions on the statistical decay, everything else being equal, we used simple standard models for all these inputs: the Gilbert-Cameron parameterization of the Fermi-Gas model for level densities with the same default parameterization as the TALYS code [22], the Kopecky-Uhl fit of

$\gamma$-strength functions [23] and the Koning-Delaroche optical model for the transmission coefficients [24]. The low-lying spectra are provided by the RIPL-3 files [25].

Note that the phenomenological results in Figure 4 of the main text cannot be taken as an exact match to the CGMF calculations reported in [26]. Most of the numerical details of these calculations are not specified in that paper, so the choices made there cannot be reproduced with certainty. In addition, CGMF implements several systematics, notably those related to the level densities, that are not available in YAHFC. Most importantly, the option to impose a cutoff on the energy of $\gamma$ rays is also not available in the current version of FETA.

---


\fi
\fi

\end{document}